\documentclass[11pt,onecolumn]{article}
\usepackage{amsfonts}
\usepackage{pifont}
\usepackage{amssymb}
\usepackage{amsmath}
\usepackage{multirow}
\usepackage{CJK}

\usepackage{caption}
\usepackage{latexsym}
\usepackage{graphicx}
\usepackage{epsfig}

\normalsize \makeatletter

\newtheorem{theorem}{Theorem}[section]

\newcommand{\head}[1]
{\markright{\hbox to 0pt{\vtop to 0pt{\hbox{}\vskip 3mm \hrule
width \textwidth \vss} \hss}{\sc #1}}}

\begin{document}
\title{Tighter Bounds on Non-clairvoyant Parallel Machine Scheduling with Prediction to Minimize Makespan \\
(Extended Abstract)}
\author{Tianqi CHEN\thanks{School of Mathematical Science, Zhejiang University, Hangzhou 310027, P. R. China.} \and Zhiyi TAN\thanks{Corresponding author. School of Mathematical Science, Zhejiang University,
Hangzhou 310027, P. R. China. Supported by the National Natural
Science Foundation of China (12071427). {\tt
tanzy@zju.edu.cn}} }
\date{}
\maketitle

\section{Introduction}

In recent years, with the rapid development of artificial intelligence and machine learning technologies, online algorithms with prediction have garnered widespread attention \cite{MV22}. The most fundamental characteristic that distinguishes online problems from offline problems is the lack of necessary information at the time of decision-making, which poses insurmountable difficulties in designing algorithms with good approximate performance. Fortunately,  obtaining information through predictive methods can alleviate this dilemma to some extent. Although it is often the case that information is incomplete in real-world situations, it is generally possible to make predictions about such missing information. However, we must recognize that predictions inevitably contain errors and should not be equated with actual values.

In this paper, we investigate non-clairvoyant parallel machine scheduling with prediction \cite{BKLP23}. Consider a set of $n$ independent jobs $\mathcal{J} =\{J_1, \ldots, J_n\}$ that are to be processed on $m\geq 2$ parallel identical machines, denoted as $M_1, M_2,\ldots, M_m$. All the jobs are available at time zero. The actual processing time for job $J_j$ is $p_j$, which is only revealed upon completion of the job. Nevertheless, the processing times of all jobs can be predicted by a machine learning oracle, and these predictions are accessible at time zero. The predicted processing time for job $J_j$ is denoted as $q_j$. Due to the uncertainty of the actual processing time before the completion of a job, the scheduling of the jobs can only be carried out over time and cannot be completed at time zero. Whenever a machine becomes idle, an unassigned job can be allocated to that machine for processing. Continue this process until all jobs are completed. If processing of a job can be preempted and later resumed possibly on a different machine, this scenario is termed the preemptive case. Otherwise, it is known as the non-preemptive case. The objective is to minimize the makespan, i.e., the maximal completion time of all the jobs.

To investigate the impact of prediction errors on algorithmic performance, we formally define the {\it error of prediction} as
$$\alpha=\max\limits_{1\leq j\leq n}\left\{\frac{p_j}{q_j},\frac{q_j}{p_j}\right \},$$
representing the maximum relative magnitude between predicted and actual processing times across all jobs. The prediction accuracy increases as $\alpha$ approaches 1. Both the lower bound of the problem and the competitive ratio of online algorithms can be expressed as explicit functions of parameter $\alpha$, illustrating the impact of prediction accuracy on the design of these algorithms.

The above problem was first studied by Bampis et al. \cite{BKLP23}. For the non-preemptive case, they demonstrated that the lower bound of the problem is at least 
$$	\left\{\begin{aligned}
	&\frac{1+\alpha^2}{2},&&\alpha\leq \sqrt{2},\\
	&1+\frac{1}{\left\lfloor \alpha^2 \right\rfloor }\left\lfloor\frac{\left\lfloor \alpha^2 \right\rfloor (m-1)}{m}\right\rfloor,\quad &&\alpha\geq \sqrt{2}.
\end{aligned}\right.
$$
They proposed an algorithm Longest Predicted Processing Time (LPPT), and proved that its competitive ratio is at most $\min\left\{\frac{2+2\alpha^2}{2},1+\frac{\alpha^2}{2}\left(1-\frac{1}{m} \right), 2-\frac{1}{m} \right\}$.  For the preemptive case, they showed the lower bound of the problem is at least $2-\frac{1}{\alpha^2}-\frac{1}{m}$, and also at least 
$$
\left\{\begin{aligned}
	&\frac{\alpha^2 m+m-1}{\alpha^2+2 (m-1)},&&\alpha\leq \sqrt{2},\\
	&2-\frac{1}{m}-\frac{m-1}{m \left\lfloor \alpha^2\right\rfloor },\quad &&\alpha\geq \sqrt{2}.
\end{aligned}\right.
$$
They proposed an algorithm Predicted Proportional Round Robin (PPRR), and proved that its competitive ratio is at most $2-\frac{\alpha^2 m+m-2}{\alpha^2 m-1}$. Note that the competitive ratio of the optimal algorithm for the non-preemptive or preemptive non-clairvoyant scheduling without prediction is $2-\frac{1}{m}$ \cite{Graham66, SWW95}. This indicates that predictive information can indeed enhance algorithm performance, particularly when the prediction error is small.

In this paper, we present improved lower and upper bounds for both the non-preemptive and preemptive cases. We will discuss the non-preemptive and preemptive cases in Sections 2 and 3, respectively.

Online parallel machine scheduling is one of the most fundamental and extensively studied online problems. For makespan minimization problems, the online over list model has been the most widely investigated. In this model, the algorithm schedules jobs in a certain order. When scheduling a job, it knows the processing time of the current job but has no knowledge of the processing times of subsequent jobs. The classical List Scheduling algorithm \cite{Graham66} is the optimal algorithm for the non-preemptive case on two and three machines. Despite numerous advancements, the optimal algorithm for four or more machines remains unknown \cite{PST04}. For the preemptive case, Chen et al. \cite{CVW95} have provided the optimal algorithm for any number of machines. 

Recent years have seen a growing interest in research on online scheduling with prediction. However, the majority of studies have concentrated on problems where the objectives are to minimize the total completion time and the total flow time \cite{PSK18, ALT21, IKQP23}. More closely related to the problem studied in this paper is the semi online scheduling with tightly grouped processing times, which falling under the online over list model with partial information. It is known at time zero that the processing time for each job lies within $[1, r]$, where $r > 1$. He and Zhang \cite{HZ99}, He and Dosa \cite{HD05} respectively provided the competitive ratios of the LS for the non-preemptive case on two and three machines with respect to $r$. The corresponding problem for the preemptive case was studied in \cite{ES11}, and analytic competitive ratios of the optimal algorithm were only obtained for two and three machines \cite{HJ04a, HJ04b, Du04}. When $\alpha$ is large, the competitive ratio of the LPPT is similar to that of the LS. However, when $\alpha$ is small, the two are clearly different. An interesting phenomenon arises when comparing the competitive ratios of the optimal algorithms for non-clairvoyant scheduling with prediction and semi online scheduling with tightly grouped processing times for the preemptive case. When $\alpha$ is small, the former is smaller, but when $\alpha$ is large, the situation reverses, at least for $m=2,3$.

\section{Non-preemptive Scheduling}

In this section, we study the problem of the non-preemptive case.     
First, we propose an improved lower bound for a general number of machines.  

\begin{theorem}\label{thm:lb}
	The competitive ratio of any algorithm for the non-preemptive non-clairvoyant  scheduling with prediction on $m$ identical machines is at least 
	$$
	\left\{\begin{aligned}
		&\frac{1+\alpha^2}{2},&&\alpha\leq \sqrt{2},\\
		&2-\frac{1}{\alpha^2},&& \sqrt{2}<\alpha< \sqrt{m},\\
		&2-\frac{1}{m},\quad&&\alpha\geq \sqrt{m}.
	\end{aligned}\right.
	$$
\end{theorem}

Next, we analyze the worst-case performance of the LPPT algorithm \cite{BKLP23}. The LPPT algorithm schedules the jobs one by one in non-increasing order of predicted processing times. Whenever a machine becomes idle, it always assigns the job with the largest predicted processing time among the unassigned jobs to that machine.
We will show a tighter competitive ratio of the LPPT for a general number of machines and tight competitive ratios for the cases of two and three machines.

\begin{theorem}\label{thm:ub}
	The competitive ratio of the LPPT for the non-preemptive non-clairvoyant  scheduling with prediction on $m$ machines is at most 
	$$
	f_n(\alpha)=\left\{\begin{aligned}
		&1+\frac{(m-1)\alpha^2}{3m},\quad&&1\leq \alpha\leq \sqrt{\frac{3m}{m+2}},\\
		&\frac{1+\alpha^2}{2},&& \sqrt{\frac{3m}{m+2}}\leq \alpha\leq \sqrt{\frac{3m-2}{m}},\\
		&2-\frac{1}{m},&&\alpha\geq \sqrt{\frac{3m-2}{m}},
	\end{aligned}\right.
	$$
and the LPPT is optimal when $\alpha\geq \sqrt{m}$.
\end{theorem}

\begin{theorem}\label{thm:non-pre_m=2}
	The competitive ratio of the LPPT for the non-preemptive non-clairvoyant  scheduling with prediction  on $2$ machines is
	$$
	f_2(\alpha)=\left\{\begin{aligned}
		&\frac{4+3\alpha^2}{4+2\alpha^2},\quad&&1\leq \alpha < \sqrt[4]{2},\\
		&\frac{1+\alpha^2}{2},&& \sqrt[4]{2}\leq\alpha < \sqrt{2},\\
		&\frac{3}{2},&&\alpha\geq\sqrt{2},
		\end{aligned}\right.
	$$
and the LPPT is optimal when $\alpha\geq \sqrt[4]{2}$.
\end{theorem}

\begin{theorem}\label{thm:non-pre_m=3}
	The competitive ratio of the LPPT for the non-preemptive non-clairvoyant  scheduling with prediction  on $3$ machines is
	$$
    f_3(\alpha)=\left\{\begin{aligned}
		&\frac{6+5\alpha^2}{6+3\alpha^2},\quad&&1\leq \alpha < \sqrt{\frac{1+\sqrt{17}}{4}},\\
		&\frac{5+2\alpha^2}{6},&& \sqrt{\frac{1+\sqrt{17}}{4}}\leq\alpha < \frac{\sqrt{6}}{2},\\
		&\frac{3+2\alpha^2}{3+\alpha^2},&&\frac{\sqrt{6}}{2}\leq \alpha < \sqrt[4]{3},\\
		&\frac{1+\alpha^2}{2},&&\sqrt[4]{3}\leq \alpha < \sqrt{2}.\\
		&2-\frac{1}{\alpha^2},&&\sqrt{2}\leq \alpha < \sqrt{3},\\
		&\frac{5}{3},&&\alpha \geq \sqrt{3},
	\end{aligned}\right.
	$$
and the LPPT is optimal when $\alpha\geq \sqrt[4]{3}$.	
\end{theorem}

\section{Preemptive Scheduling}

In this section, we study the problem of the preemptive case. For the preemptive scheduling, the processing time of a job can be interpreted as its total processing requirement. Processing a job for a duration $\delta$ on any machine reduces its remaining processing requirement by $\delta$. However, although job processing can be preempted, for any schedule generated by an algorithm, the total processing requirement processed by all machines at time $t$ never exceeds $mt$. This fundamental constraint inevitably implies that at least one job has been processed for no more than $\frac{mt}{n}$ of its processing requirement. Based on the above arguments, we provide an improved lower bound for the preemptive scheduling.

\begin{theorem}\label{lm:prelb1}
	The competitive ratio of any algorithm for the preemptive non-clairvoyant  scheduling with prediction on $m$ machines is at least 
$$\max\left\{2-\frac1{m}+\left(\frac1m-1\right)\frac{m-1}{\lfloor(m-1)\alpha^2\rfloor},\frac{\lceil(m-1)\alpha^2\rceil +m\alpha^2-m+1}{\alpha^2+\lceil(m-1)\alpha^2\rceil}\right\}.$$	 
\end{theorem}

Next, we propose a tighter competitive ratio for the algorithm PPRR. This algorithm employs the standard alternating processing technique in preemptive scheduling. Under preemptive scheduling constraints, a single job cannot be processed simultaneously on two machines, while multiple jobs may be alternately processed on different machines at varying speeds. More formally,  $J_{j_1}, J_{j_2}, \dots, J_{j_h}$ are alternately processed on $g\leq h$ machines with speed $s_{j_i}(t)$ at time $t$, where $0\leq s_{j_i}(t)\leq 1$ and $\sum_{i=1}^h s_{j_i}(t)=g$, implies that during an infinitesimally small time interval $[t,t+\delta)$, the remaining processing requirement of $J_{j_i}$ decreases by $ s_{j_i}(t)\delta$, $i=1,\ldots,h$.  

{\bf Algorithm PPRR} (\cite{BKLP23})

Initialize all machines as unoccupied and all jobs as non-mandatory. The following procedure is executed whenever the system initializes or when one or more jobs complete. 

Let there be $g$ unoccupied machines and $h$ uncompleted non-mandatory jobs $J_{j_1}, J_{j_2}, \dots, J_{j_h}$ with $q_{j_1}\geq q_{j_2}\geq \dots \geq q_{j_h}$ at this moment. 

1. Determine
$$d=\max\left\{k| (g-k)q_{j_k}\geq \sum_{i=k+1}^h q_{j_i} {\rm \ and \ } 1\leq k\leq \min\{g,h \}\right \}.$$
Jobs $J_{j_1}, J_{j_2},\ldots, J_{j_d}$ are designated as mandatory jobs. 

2. Select $d$ machines arbitrarily from the $g$ unoccupied machines to become occupied machines. Each selected machine processes a single job from $J_{j_1}, J_{j_2},\ldots, J_{j_d}$ until completion, after which the machine returns to unoccupied again. 

3. The remaining jobs $J_{j_{d+1}}, J_{j_{d+2}},\ldots, J_{j_{h}}$ are alternately processed on the remaining $g-d$ unoccupied machines with the speed of $J_{j_{i}}$ given by $\frac{(g-d)q_{j_{i}}}{\sum_{i=d+1}^l q_{j_{i}}}$, $i=d+1,\dots,h$.

\begin{theorem}\label{lm:preub1}
	The competitive ratio of the PPRR for the preemptive non-clairvoyant  scheduling with prediction on $m$ machines is at most $2-\frac{1}{m}-\frac{m-1}{m \alpha^2}$, and the PPRR is optimal when $(m-1)\alpha^2$ is an integer.	 
\end{theorem}

\end{document}